\documentclass[preprint,showpacs,superscriptaddress,prd]{revtex4}
\usepackage{graphicx,amsmath,dcolumn,bm}
\begin{document}
\title{\Large{\bf{Nuclear effects on $J/\psi$ production in proton-nucleus collisions}}}
\author{ Chun-Gui  Duan}
\affiliation{Department of Physics, Hebei Normal
             University,
             Shijiazhuang 050016, P.R.China}
\affiliation{  Hebei Advanced Thin Films Laboratory, Shijiazhuang
               050016, P.R.China}
\author{Jian-Chao  Xu}
\affiliation{Department of Physics, Hebei Normal
             University,
             Shijiazhuang 050016, P.R.China}

\author{Li-Hua Song}
\affiliation{Department of Physics, Hebei Normal
             University,
             Shijiazhuang 050016, P.R.China}
\affiliation{ College of information,  Hebei Polytechnic University,
Tangshan 063009, P.R.China}


\begin{abstract}

The study of nuclear effects for $J/\psi$ production in
proton-nucleus collisions is  crucial for a correct interpretation
of the $J/\psi$ suppression patterns experimentally observed in
heavy-ion collisions. By means of three representative sets of
nuclear parton distributions, the energy loss effect in initial
state and nuclear absorption effect in final state are taken into
account in the uniform framework of the Glauber model. A leading
order phenomenological analysis is performed on $J/\psi$ production
cross section ratios $R_{W/Be}(x_{F})$ for the E866 experimental
data. The $J/\psi$ suppression is investigated quantitatively due to
the different nuclear effects.  It is shown that the energy loss
effect with resulting in the suppression on $R_{W/Be}(x_{F})$ is
more important  than the nuclear effects on parton distributions in
high $x_{F}$ region. The E866 data in the small $x_{F}$ keep out the
nuclear gluon distribution with a large anti-shadowing effect.
However, the new HERA-B measurement is not in support of the
anti-shadowing effect in nuclear gluon distribution. It is found
that the $J/\psi$-nucleon inelastic cross section
$\sigma^{J/\psi}_{abs}$ depends on the kinematical variable $x_{F}$,
and increases as $x_{F}$ in the region $x_{F}>0.2 $.

\end{abstract}


\maketitle
\newpage
\vskip 0.5cm

\section{Introduction}

The production of $J/\psi$ in high energy collisions has attracted
the extensive attention from both the nuclear and particle physics
communities$^{[1]}$.  $J/\psi$ suppression  is considered as a most
reliable signature for the formation of  Quark Gluon Plasma (QGP) in
heavy ion collisions during the present time.   However, there are
many different nuclear effects in nucleus-nucleus collisions.  A
robust interpretation of the experimental data in heavy ion
collisions requires to understand deeply and quantitatively the
basic mechanisms responsible for the suppression of  $J/\psi$
production due to the nuclear effects.  Therefore,  it is desirable
that a good baseline should be established by means of the study on
$J/\psi$ production in proton-nucleus collisions to clarify the
conventional nuclear suppression mechanism.

Several proton-induced fixed target experiments(such as,
NA3$^{[2]}$, E772$^{[3]}$, E866$^{[4]}$, NA50$^{[5]}$  and
HEAR-B$^{[6]}$) have studied the nuclear dependence of $J/\psi$
production cross-sections as a function of the Feynman variable
$x_{F}$.    These experimental measurements employ the usual
parametrization by a power law: $\sigma_{pA} =\sigma_{pN} \cdot
A^\alpha$,  where $\sigma_{pN}$ is the proton-nucleon cross section
and $\sigma_{pA}$ is corresponding proton-nucleus cross section for
a target of atomic mass number $A$. The E866 collaboration$^{[4]}$
published the precise measurement of the suppression factor $\alpha$
for 800 GeV protons incident on  iron and tungsten nuclear targets,
relative to  beryllium  nuclear targets with very broad coverage
$-0.1 <x_F < 0.95$.   The observed suppression is smallest at
$x_{\mathrm{F}}$ values of 0.25 and below, and increases at larger
values of $x_{F}$. Recently, the HERA-B collaboration$^{[6]}$
reported the first measurement of nuclear effects in $J/\psi$
production extending into the larger negative part of the
Feynman-$x$ spectrum, $-0.34 <x_F < 0.14$ by 920GeV protons incident
on carbon, titanium and tungsten targets. The E866 and HERA-B
results are compatible within statistical and systematic
uncertainties in the overlap region, and are systematically above
the NA50 results$^{[5]}$ with based on lower energy collisions.   At
lower values of  $x_F$,  the HERA-B $\alpha(x_{F})$ measurement
indicates a reversal of the suppression trend seen at high $x_F$:
the strong suppression established by previous measurements at high
$x_{F}$ turns into a slight tendency towards enhancement in the
negative $x_{F}$ region.

It is now clear that $J/\psi$ production occurs in two different
steps where a charm quark pair is produced first through the
interaction of a projectile on a target parton,  followed by the
non-perturbative formation of the colorless asymptotic state.  As a
consequence, color octet as well as singlet $c\bar{c}$ states
contribute to $J/\psi$ production.  Two formalisms have been
proposed to incorporate these features, which are  the
non-relativistic QCD (NRQCD) $^{[7]}$, and the color evaporation
model (CEM) $^{[8]}$, respectively. The CEM shows some similarities
with NRQCD.  In CEM calculations, $J/\psi$ production via
color-octet processes is allowed, hence the kinematic dependence of
the cross section is similar to NRQCD in the production of $J/\psi$.
The CEM  and NRQCD predictions have been successful in charmonium
phenomenology. Therefore, the CEM and NRQCD are used to date as a
model for charmonium production. It  is interesting to notice that
the measured $J/\psi$ polarization favors the CEM and is not within
the predictions of NRQCD,  and the CEM has fewer free parameters
than NRQCD$^{[9]}$.

The suppression of $J/\psi$ production in proton-nucleus collisions
could be caused by many different nuclear effects.   The nuclear
effects are usually subdivided in "initial state effects" related to
the projectiles which produce a charm quark pair  and "final state
effects" from interactions of the charm quark pair or the fully
formation of a $J/\psi$ in the nuclear environment.  The nuclear
effects on parton distribution functions and initial state energy
loss effect are considered as the most important initial state
effects,  while the so-called  nuclear absorption of the charm quark
pair is the main final state effect.

After the production of a charm quark pair,  the pre-meson charm
quark pair will travel through the nuclear environment. They are
subject to the strong interactions with the nuclear matter. Such
interactions,    which reduce the probability that the pre-meson
state will ultimately form a  $J/\psi$ without being absorbed,  can
be described by the Glauber model$^{[10]}$. The nuclear absorption
cross section  is introduced to express the $J/\psi$ production in
proton-nucleus collisions.

After the discovery of the EMC effect$^{[11]}$, it is well known
that nuclear  parton distribution functions are  different from
those in a free nucleon.  The nuclear modifications relative to the
nucleon parton distribution functions, are usually referred to as
the nuclear effects on the parton distribution functions, which
include nuclear shadowing, anti-shadowing, EMC effect and Fermi
motion effect in different regions of parton momentum fraction. As
for the nuclear parton distributions, the global analysis method has
been proposed in the recent years. So far, three groups have
presented global  analysis of the nuclear parton distribution
functions analogous to those of the free proton. These are the ones
by Eskola et al. ( EKS98 $^{[12]}$ and EPS$^{[13]}$), by Hirai et
al. (HKM$^{[14]}$, HKN04$^{[15]}$ and HKN07$^{[16]}$), and by de
Florian and Sassot (nDS) $^{[17]}$.  It is noticeable that the
nuclear valence quark distributions from different global analysis
are nicely in good agreement. The nuclear sea quark and gluon
distributions in general are still quite  badly constrained,
resulting in large differences between the different sets.

The initial state energy loss effect is another nuclear effect apart
from the nuclear effects on the parton distribution as in deep
inelastic scattering. In high energy proton-nucleus scattering, the
projectile rarely retains a major fraction of its momentum in
traversing the nucleus. The  projectile can lose a finite fraction
of its energy  due to the  multiple collisions and repeated energy
loss in the nuclear target.  The proton incident Drell-Yan
reaction$^{[18]}$ on nuclear targets provides, in particular, the
possibility of probing the propagation of projectile through nuclear
matter, with the produced lepton pair  not interacting strongly with
the partons in the nuclei. Therefore, the nuclear Drell-Yan process
is an ideal tool to study initial state energy loss effect. At
parton level, the nuclear Drell-Yan process is only sensitive to the
quark energy loss$^{[19]}$. However, a charm quark pair is
predominantly due to gluon fusion in proton induced $J/\psi$
production on nuclear targets.  The gluon energy loss can not yet be
constrained by the experimental data though  some theoretical
papers$^{[20]}$ on $J/\psi$ production employ the conclusion that
the  gluon energy loss is 9/4 times larger than the quark energy
loss due to the difference in the color factors. As a result, the
initial state energy loss effect is not pinned down in $J/\psi$
production which is predominantly due to gluon fusion.

In the previous article$^{[21]}$, by using the nuclear parton
distribution functions from the global analysis, the nuclear
Drell-Yan production cross section ratios were  investigated for
800GeV protons incident on a variety of nuclear targets in the
framework of Glauber model. At hadron level, the energy loss of the
beam proton in nuclear environment was determined well by fitting
the nuclear Drell-Yan data from the Fermilab E866
experiment$^{[22]}$. In this paper, by combining three
representative sets of nuclear parton distribution functions with
the initial state energy loss of the incident proton determined by
the nuclear Drell-Yan reaction, $J/\psi$ production cross section
ratios for 800GeV protons incident on tungsten and beryllium nuclear
targets are analyzed  with meanwhile taking  account of the nuclear
absorption effect in the Glauber model.  It is hoped to have a new
knowledge about the nuclear effects on $J/\psi$ production in
proton-nucleus collisions.

The paper is organized as follows. In sect.II, a brief formalism for
the  differential cross section in $J/\psi$ production is presented.
The section III is our results and discussion. The summary is given
in sect.IV.

\section{The formalism for $J/\psi$
production differential cross section }

At leading order(LO) in perturbative QCD, the charmonium production
cross section is  the sum of two partonic contributions, gluon
fusion ($gg$) and quark-anti-quark annihilation ($q\bar{q}$),
convoluted with the  parton distribution functions in the incident
proton  $p$ and the target nucleus $A$$^{[23]}$:
\begin{eqnarray}
  \frac{d\sigma}{d x_{_{\rm F}}} &=& \rho_{_{J/\psi}}
  \int_{2 m_{_c}}^{2 m_{_D}} d{m} \ \frac{2 m}{\sqrt{x_{_{\rm F}}^2 s + 4 m^2}}
  \ \Bigg[ f_g^{p}(x_{_1},m^2) \ f_g^{A}(x_{_2},m^2) \  \sigma_{gg}(m^2) \nonumber \\
&+& \sum_{q=u,d,s} \bigg\{ f_q^{p}(x_{_1},m^2) \
f_{\bar{q}}^{A}(x_{_2},m^2)  + f_{\bar{q}}^{p}(x_{_1},m^2) \
f_q^{A}(x_{_2},m^2) \bigg\} \sigma_{q\bar{q}}(m^2) \Bigg],
\end{eqnarray}
where $x_{1(2)}$ is the projectile proton ( target) parton momentum
fractions, $x_F=x_1-x_2$,  $\sqrt{s}$ is the center of mass energy
of the hadronic collision, $m^2=x_1x_2 s $,  $m_{_c}=1.2$GeV and
$m_{_D}=1.87$GeV are respectively the charm-quark  and $D$ meson
mass, and $\sigma_{gg}$ ( $\sigma_{q\bar{q}}$) is the LO $c\bar{c}$
partonic production cross section from the gluon
fusion(quark-antiquark annihilation). $\rho_{_{J/\psi}}$ is the
fraction of $c\bar{c}$ pair which produces the $J/\psi$ state,
$f_{i}^{p}$ and $f_{i}^{A}$ stand respectively for the parton
distribution function in the proton and in the nucleus.

According to Glauber model$^{[10]}$,  the projectile proton
scattering inelastically on nucleus (A) makes many collisions with
nucleons bound in  nuclei. The probability of having $n$ collisions
can be expressed as
\begin{equation}
P(n)=\frac{\int d\vec{b}P(n,\vec{b})}{\sum\limits_{n=1}^{A}\int
d\vec{b}P(n,\vec{b})},
\end{equation}
where
\begin{equation*}
P(n,\vec{b})=\frac{A!}{n!(A-n)!}[T(\vec{b})\sigma_{in}]^{n}[1-
T(\vec{b})\sigma_{in}]^{A-n},
\end{equation*}
$\sigma_{in}$($\sim 30mb$) is non-diffractive cross section for
inelastic nucleon-nucleon collision, and $T(\vec{b})$ is the
thickness function of impact parameter $\vec{b}$.

Now let us take  account of the energy loss of the projectile proton
moving through the target nuclei before producing  the charm quark
pair in proton-nucleus. The energy loss due to beam proton can
induce the decrease of center of mass system energy of the
nucleon-nucleon collision producing  $c\bar{c}$, and affect the
measured $J/\psi$ production cross section. After the projectile
proton has $n$ collisions with nucleons in  nuclei,  it is supposed
for convenient calculation that the center-of-mass system energy of
the nucleon-nucleon collision  can be expressed as
\begin{equation}
\sqrt{s^{\prime}}=\sqrt{s}-(n-1)\triangle\sqrt{s},
\end{equation}
where $\triangle\sqrt{s}$ is the center-of-mass system  energy loss
per collision in the initial state(see Ref.[21] for more detail
discussion). Therefore, the $J/\psi$ production cross section in the
nth collision can be rewritten as
\begin{equation}
\frac{d\sigma^{(n)}}{d x_{_{\rm F}}}=\frac{d\sigma}{d
x^{\prime}_{_{\rm F}}}.
\end {equation}
Here the rescaled quantities are defined as
\begin{equation}
x_{F}^{\prime}=r_{s}x_{F}, \hspace{1cm}
x_{1,2}^{\prime}=r_{s}x_{1,2},
\end{equation}
with the centre-of-mass system energy ratio:
\begin{equation}
r_{s}=\frac{\sqrt{s}}{\sqrt{s^{\prime}}}.
\end{equation}

Furthermore, the nuclear absorption of the charm quark pair is added
in final state effect. In the framework of the Glauber model, the
probability for no interaction (or ``survival probability'') of the
$J/\psi$ meson with the target nucleus$^{[24]}$, can be calculated
as
\begin{equation}
S_{\rm{abs}} = \frac{1}{(A-1)\sigma^{J/\psi}_{abs}}\int
d\vec{b}(1-e^{-(A-1)T(\vec{b}) \sigma^{J/\psi}_{abs}}).
\end{equation}
The survival probability depends obviously on both the atomic mass
number $A$ of the target nucleus and the $J/\psi$-nucleon inelastic
cross section $\sigma^{J/\psi}_{abs}$. If the factorization between
the $c\bar{c}$ production process and the subsequent possible
$J/\psi$ inelastic interaction with nuclear matter is assumed,  the
$J/\psi$ production cross section in the nth collision is written as
\begin{equation}
\frac{d\sigma^{(n)}}{d x_{_{\rm F}}}=\frac{d\sigma}{d
x^{\prime}_{_{\rm F}}}S_{\rm{abs}}.
\end{equation}

Combining above ingredients on initial  and final state effects, the
$J/\psi$ production  cross section in proton-nucleus collisions can
be expressed as
\begin{equation}
\langle \frac{d\sigma}{d x_{_{\rm
F}}}\rangle=\sum\limits_{n=1}^{A}P(n)\frac{d\sigma^{(n)}}{d x_{_{\rm
F}}}.
\end{equation}

\section{ Results  and discussion }

The Fermilab Experiment866(E866) $^{[4]}$  measured the differential
cross section ratios of proton induced tungsten to beryllium target
for $J/\psi$ production,
\begin{equation}
R_{W/Be}(x_{F})=\frac{\langle \frac{d\sigma^{p-W}}{d x_{_{\rm
F}}}\rangle}{\langle \frac{d\sigma^{p-Be}}{d x_{_{\rm F}}}\rangle}.
\end{equation}
The covered kinematical range was $-0.1 <x_F < 0.95$. The
experimental data were provided in  small $x_{F}$ ( $-0.1 <x_F <
0.3$), intermediate $x_{F}$ ( $0.2 <x_F < 0.65$), and large $x_{F}$
( $0.3 <x_F < 0.95$), respectively. By using three sets of leading
order nuclear parton distribution functions together with CTEQ6L
parton density in the proton$^{[25]}$, meanwhile taking account of
the energy loss of the beam proton in initial state effect and the
nuclear absorption of the charmonium states traversing the nuclear
matter in the uniform framework of the Glauber model, a leading
order phenomenological analysis is given in the color evaporation
model to the E866 experimental data on the ratios of $J/\psi$
production differential cross section $R_{W/Be}(x_{F})$.

As for the nuclear parton distributions in our calculation,  the
nuclear modification for gluon distribution function is apparently
different between the different sets. The Figure 1 shows the ratio
of the tungsten nucleus over the proton parton distribution function
at the charm quark pair mass scale, $R_g^{\rm W}(x, Q^2=(2m_c)^2)$,
as a function of Bjorken $x$ for the leading order nDS (dotted
line), EKS98 (solid line) and HKN07 (dashed line) nuclear gluon
distribution functions. The main differences between the three
nuclear parameterizations are found that nDS and EKS98 nuclear gluon
densities have respectively nuclear anti-shadowing effect in the
region $0.03 <x < 0.25$ and $0.03 <x < 0.38$. It is emphasized  that
nDS nuclear gluon density have comparatively very small
anti-shadowing than EKS98 in intermediate $x$ region, whereas HKN07
nuclear gluon distribution function has only the nuclear shadowing
effect in small $x$. The ratio $R_g^{\rm W}$ from HKN07 increases as
$x$ becomes larger, and crosses the line $R_g^{\rm W}=1$ at $x\sim
0.1$.

\begin{figure}[!]
\centering
\includegraphics*[width=90.3mm,height=9cm]{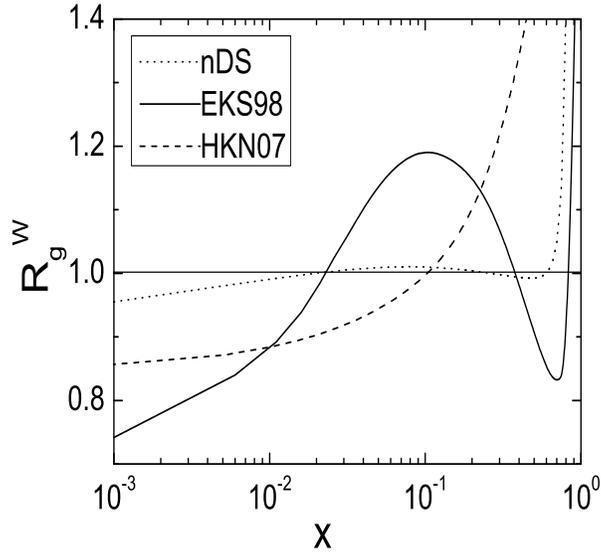}
\vspace{-0.25cm} \caption{The ratio $R_g^{\rm W}$ of the gluon
distribution in a tungsten nucleus over that in a free proton.  The
solid, dotted and dashed lines correspond to the results given by
the EKS98, nDS and HKN07 nuclear parton distributions,
respectively.}
\end{figure}

If neglecting the energy loss  in initial state effect and the
nuclear absorption in final state effect, the calculated results are
compared with the E866 experimental data$^{[4]}$ on
 $J/\psi$ production cross section ratios $R_{W/Be}(x_{F})$ in Fig.2.  The solid, dotted and
dashed lines are the results on $R_{W/Be}(x_{F})$  by using the
EKS98, nDS and HKN07 parameterizations, respectively.  As can be
seen in Fig.2, $R_{W/Be}(x_{F})$ from  nDS nuclear parton
distributions shows a remarkably flat behavior as a function $x_{F}$
in the region $|x_{F}|<0.1 $. However, $R_{W/Be}(x_{F})$ by HKN07
parameterizations turns into a tendency towards enhancement in the
negative $x_{F}$ region.  The anti-shadowing enhancement is appeared
in the EKS98 model at $x_{\rm F} \approx 0$. It is worth noting that
the nuclear suppression from the nuclear effects on the parton
distributions  becomes larger as the increase of $x_{F}$ in the
range $x_{F}>0 $. With taking example by HKN07 nuclear parton
distribution functions, the suppression is approximately $4\%$ to
$14\%$ for $R_{W/Be}(x_{F})$ in the ranges $0.1\leq x_F\leq 0.95$.

\begin{figure}[t,m,b]
\centering
\includegraphics*[width=90.3mm,height=9cm]{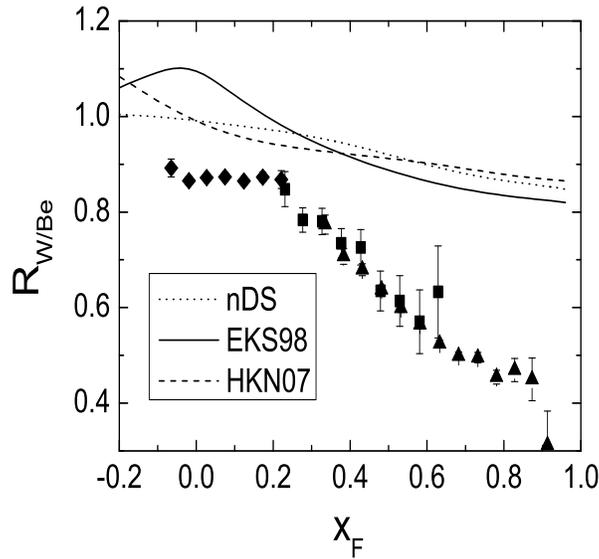}
\vspace{-0.25cm} \caption{The  $J/\psi$ production cross section
ratio $R_{W/Be}(x_{F})$.  The filled diamonds (
   boxes and triangles) are the E866 experimental data $^{[4]}$
in  the region $-0.1 <x_F < 0.3$  ( $0.2 <x_F < 0.65$ and  $0.3 <x_F
< 0.95$). With assuming only
  the nuclear effects on the parton distribution functions,  the solid, dotted
and dashed lines correspond to the results from  the EKS98, nDS and
HKN07 parameterizations, respectively.}
\end{figure}

The curves in Fig.3 show  the  calculated $R_{W/Be}(x_{F})$ for
$J/\psi$ production, compared with the E866 data, given by using the
EKS98, nDS and HKN07 nuclear parton distributions together with the
energy loss of the beam proton in initial state effects. In our
calculations, the center-of-mass system  energy loss per collision
$\triangle\sqrt{s}=$  0.18GeV  was determined from the nuclear
Drell-Yan experimental data in the Glauber model$^{[21]}$. It is
found that the energy loss effect of the beam proton in initial
state has a small impact on the differential cross section ratios
$R_{W/Be}(x_{F})$ in the region $x_{F}<0.2$. However, the energy
loss effect adds further the nuclear suppression on
$R_{W/Be}(x_{F})$ in the range $x_{F}>0.2$. The nuclear suppression
on $R_{W/Be}(x_{F})$ from energy loss effect increases gradually in
the region $0.2\leq x_F\leq 0.8$, and becomes much steeper in the
region $x_{F}>0.8$. The suppression due to energy loss effect is
approximately $1\%$ to $14\%$ and $14\%$ to $50\%$ in the ranges
$0.2\leq x_F\leq0.8$ and $0.8\leq x_F\leq0.96$, respectively. As for
HKN07 nuclear parton distributions, the total suppression from the
nuclear effects on parton distribution functions and energy loss
effect is roughly $4\%$ to $27\%$  and $27\%$ to $65\%$ in the
ranges $0.2\leq x_F\leq0.8$ and $0.8\leq x_F\leq0.96$, respectively.
The similar results can be obtained from nDS and EKS98 nuclear
parton distributions. Therefore, the energy loss effect, resulting
in the suppression on $R_{W/Be}(x_{F})$, is more important  than the
nuclear effects on parton distributions in high $x_{F}$ region.
Although the initial state effects on $R_{W/Be}(x_{F})$ become
larger as the increase of $x_{F}$, especially in high $x_{F}$
region, the remained deviation from the E866 data need to be
contributed by the final state effect.

\begin{figure}[t,m,b]
\centering
\includegraphics*[width=90.3mm,height=9cm]{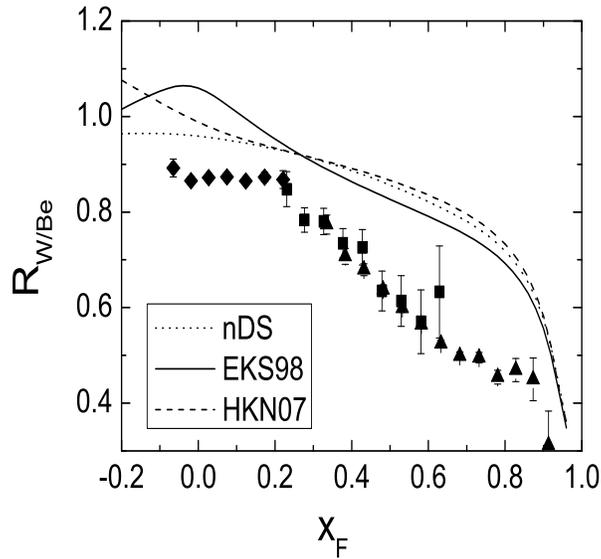}
\vspace{-0.25cm} \caption{The ratio $R_{W/Be}(x_{F})$ by combining
the nuclear effects on the parton distributions with the energy loss
effect in initial state. The  comments are the same as Fig.2.}
\end{figure}

After considering the initial state effects, it is supposed further
that the nuclear absorption of the charm quark pair is the main
final state effect.  The ratios of $J/\psi$ production differential
cross section $R_{W/Be}(x_{F})$ are calculated in the small $x_{F}$
region by combining the nuclear effects on parton distributions with
the energy loss and nuclear absorption effect in the uniform
framework of the Glauber model. The theoretical results are shown in
Fig.4 against the E866 experimental data on $R_{W/Be}(x_{F})$. The
solid, dotted and dashed lines stand for  the results from the
EKS98, nDS and HKN07 parameterizations, respectively. The
$J/\psi$-nucleon inelastic cross section $\sigma^{J/\psi}_{abs}$ (in
unit mb)and the $\chi^2$ per degrees of freedom are calculated and
summarized in Table I by fitting the experimental data.

\begin{table}[t,m,b]
\caption{The $J/\psi$-nucleon inelastic cross section
$\sigma^{J/\psi}_{abs}$ and $\chi^2/d.o.f.$ in the small
$x_{\mathrm{F}}$ region.}
\begin{ruledtabular}
\begin{tabular*}{\hsize}
{c@{\extracolsep{0ptplus1fil}}c@{\extracolsep{0ptplus1fil}}
c@{\extracolsep{0ptplus1fil}} c@{\extracolsep{0ptplus1fil}}}
   & HKN07 & nDS & EKS98  \\
\colrule
$\sigma^{J/\psi}_{abs}(mb)$      & 2.30 & 2.01&3.60   \\
 \colrule
 $\chi^2/d.o.f.$      & 4.371 & 1.066&11.758   \\
\end{tabular*}
\end{ruledtabular}
\end{table}
\begin{figure}[!]
\centering
\includegraphics*[width=90.3mm, height=9cm]{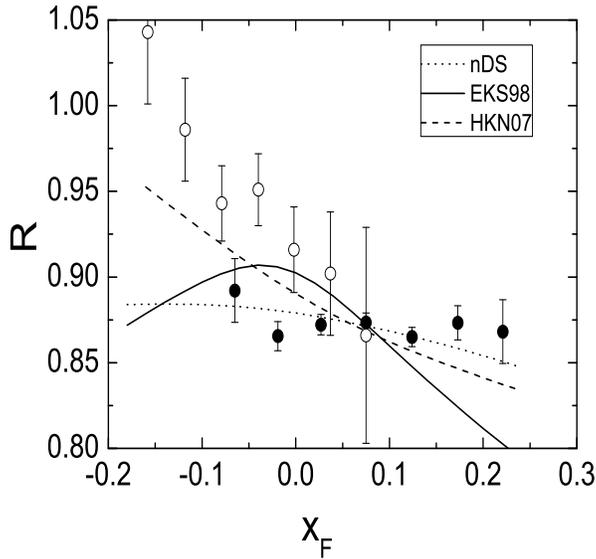}
\vspace{-0.5cm} \caption{The  $J/\psi$ cross section ratio
$R(x_{F})$. The solid circles are the ratios on tungsten to
beryllium target from the E866$^{[4]}$. The open circles are the
ratios on tungsten to carbon target from the HERA-B
experiment$^{[6]}$. The solid, dotted and dashed lines are the
calculated $R_{W/Be}(x_{F})$ from  the EKS98, nDS and HKN07
parameterizations by considering the initial state effects and final
state nuclear absorption effect with a constant
$\sigma^{J/\psi}_{abs}$.}
\end{figure}

As can be seen in Fig.4, the E866 experimental data points show
remarkably flat behavior as a function of $x_F$ in small
$x_{\mathrm{F}}$. The covered kinematical range $-0.1 <x_F < 0.3$
correspond roughly to the parton momentum fraction $0.02 <x< 0.12$
in target nuclei.  The target parton momentum fraction increases as
the decrease of $x_{\mathrm{F}}$ in the negative $x_{F}$ region.  In
the range $-0.1 <x_F < 0.3$, the gluon fusion channel dominates over
the $q\bar{q}$ annihilation process in $J/\psi$ production.  By
compared this observation with the predictions given by the
different nuclear parton distribution functions, the tendency
reported by the E866 data is well reproduced by the nDS and HKN07
parameterizations in small $x_{\mathrm{F}}$ domain. These two
parameterizations do not show a strong anti-shadowing effect,
whereas the strong anti-shadowing presents in EKS98 nuclear gluon
distribution at $x \sim 0.1$. It is shown that the E866 experimental
data consequently tend to favor nDS and HKN07 sets rather than the
EKS98 parameterizations. Recently, the HERA-B measurement$^{[6]}$
indicates that the nuclear suppression on $J/\psi$ production cross
section ratios turns into a slight tendency towards enhancement in
the negative $x_{F}$ region. In Fig.4, the open circles show the
ratios $R_{W/C}(x_{F})$ (the Table 3 in Ref.[27])which are derived
from the HERA-B $\alpha(x_{F})$ measurement(the Table 5.1 in
Ref.[6]). It is apparent that the enhancement tend is only produced
by HKN07 parameterizations, which do not present any anti-shadowing
effect. Therefore, it is concluded that the current experimental
data is not in support of the anti-shadowing effect in nuclear gluon
distribution. Our analysis shows similar tendency with the results
in Ref.[26] in the anti-shadowing part by considering E866
measurement, but the new HERA-B data further rule out nDS nuclear
gluon distribution with weak anti-shadowing effect.  The
paper$^{[26]}$ used the color evaporation model, and investigated
$J/\psi$ production cross section ratio $R_{W/Be}(x)$ as a function
of the target momentum-fraction $x(0.02<x<0.1)$ from E866$^{[4]}$ by
using the various nuclear parton distributions and keeping
$\sigma^{J/\psi}_{abs}$ as a free parameter. Their result without
considering HERA-B measurement is in favor of the nuclear gluon
distributions which do not predict a strong anti-shadowing at $x
\sim 0.1$ nor a large shadowing at small $x$. In detail, the E866
data agree with nDS and HKM $^{[14]}$(without nuclear gluon
anti-shadowing effect similar to HKN07)parametrizations, rather than
the EKS98 nuclear gluon distribution. In addition, the
$\sigma^{J/\psi}_{abs}$ is slight bigger than our results in Table I
because of  not including the energy loss effect in initial state.

\begin{figure}[!]
\centering
\includegraphics*[width=90.3mm, height=9cm]{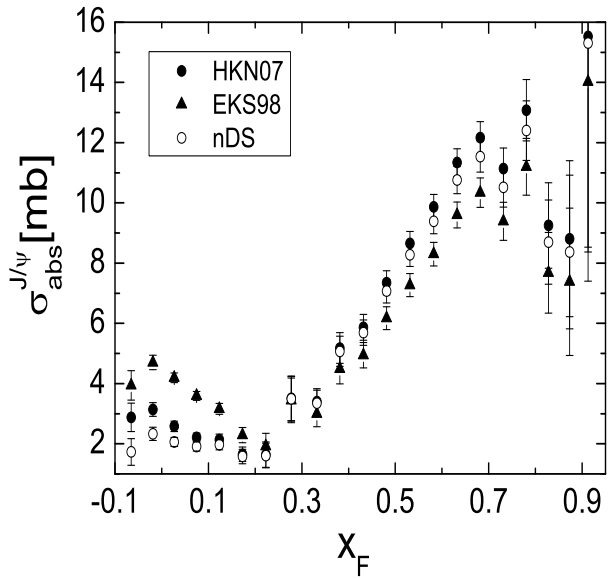}
\vspace{-0.5cm} \caption{The $J/\psi$-nucleon inelastic cross
section $\sigma^{J/\psi}_{abs}$ as a function of $x_{F}$ by means of
HKN07(solid circles), EKS98(filled triangles) and nDS(empty circles)
nuclear parton distribution functions. }
\end{figure}

The calculations on $R_{W/Be}(x_{F})$ extended to the full
experimental range in  $x_{F}$ indicate that a single value of
$\sigma^{J/\psi}_{abs}$ does not provide a good agreement with the
E866 experimental data.  Therefore, the $J/\psi$-nucleon inelastic
cross section $\sigma^{J/\psi}_{abs}$ is obtained as a function of
$x_{F}$ by fitting the experimental data on $\alpha$ versus $x_{F}$.
In Fig.5, The solid circles, filled triangles and empty circles
correspond to the values of $\sigma^{J/\psi}_{abs}$ with based on
three different nuclear parton distribution functions. In the region
$0<x_{F}<0.2 $, the values of $\sigma^{J/\psi}_{abs}$ given by nDS
are roughly a constant because of the flat nuclear modification on
gluon distribution function. However, the calculated
$\sigma^{J/\psi}_{abs}$ from EKS98 have a steeper increase than
those from HKN07 as the decrease of $x_{F}$,  which originates from
the nuclear modification on gluon distribution by EKS98 has sharply
increase than that by HKN07 parameterizations in the corresponding
region of the target parton momentum fraction. In the region
$x_{F}>0.2 $, the value of $\sigma^{J/\psi}_{abs}$ comes to be
larger as the increase of $x_F$ if excluding the two values with
deviating from the tendency in the range  $x_{F}>0.7 $. With
difference from the conclusion in Ref.[27], the
$\sigma^{J/\psi}_{abs}$ predicted here is smaller than that without
the energy loss effect in initial state in high $x_{F}$ region. The
results for $\sigma^{J/\psi}_{abs}$ from  HKN07 nuclear parton
distributions can be represented for convenience by the simple
parametrization:
$\sigma^{J/\psi}_{abs}(x_F)=2.685-9.038x_F+36.217x^2_F $. Therefore,
it can be confirmed that $\sigma^{J/\psi}_{abs}$ depends on the
kinematical variable $x_{F}$. The neglect of energy loss effect in
$J/\psi$ suppression studies is insufficient to gain an insight into
the nuclear effects in $J/\psi$ production.

\section{Summary }

The precise identification of $J/\psi$ suppression mechanism as
signatures of quark-gluon plasma formation requires a detailed and
quantitative understanding of the nuclear effects already  present
in proton-nucleus collisions. The complex nuclear effects include
mainly the nuclear effects of parton distribution functions and
energy loss effect in initial state  together with  the nuclear
absorption of the charm quark pair in final state.  Although the
nuclear effects on the parton distribution functions have been
investigated by the global analysis analogous to those of the free
proton, the nuclear gluon distributions have large differences
between the different sets. By using the Glauber model,  the energy
loss of the beam proton in nuclear environment was determined well
by fitting the nuclear Drell-Yan data from the Fermilab E866
experiment for 800GeV protons incident on a variety of nuclear
targets$^{[21]}$. The nuclear absorption effect remained in final
state can be studied with $J/\psi$ production in 800GeV protons
incident on nuclear targets. We perform a leading order
phenomenological analysis on $J/\psi$ production cross section
ratios $R_{W/Be}(x_{F})$, and compare with the E866 experimental
data. The different nuclear effects are obtained on the $J/\psi$
suppression. It is shown that the energy loss effect in $J/\psi$
suppression  is more important  than the nuclear effects on parton
distributions in high $x_{F}$ region. In the small $x_{F}$ ($-0.1
<x_F < 0.3$) range,  the E866 data are in agreement with nDS and
HKN07 rather than the EKS98 parameterizations which give a large
anti-shadowing effect. However, the new HERA-B measurement do not
support  the anti-shadowing effect in nuclear gluon distribution,
which further rules out nDS nuclear gluon distribution. It is found
that the $J/\psi$-nucleon inelastic cross section
$\sigma^{J/\psi}_{abs}$ depends on the kinematical variable $x_{F}$,
and increases as $x_{F}$ in the region $x_{F}>0.2 $. Of course, the
precise experimental data on $J/\psi$ production will be mandatory
in proton-nucleus collisions, especially in the negative $x_{F}$
region. Therefore, it is desirable to operate precise measurements
at J-PARC$^{[28]}$ and Fermilab E906$^{[29]}$ in the future.

\vskip 1cm
{\bf Acknowledgments}
This work was supported in part by the National Natural Science
Foundation of China(10575028) and  Natural Science Foundation of
Hebei Province(A2008000137).

\end{document}